\documentclass[amsmath,amssymb,aps,eqsecnum,graphicx]{revtex4}
\usepackage[dvips]{graphicx}
\usepackage{dsfont}
\usepackage{theorem}
\usepackage{txfonts}    

\newcommand{\pt}[1]{\left( #1 \right)}                   
\newcommand{\pq}[1]{\left[ #1 \right]}                   
\newcommand{\pg}[1]{\left\{ #1 \right\}}                 
\renewcommand{\geq}{\geqslant}                           
\newcommand{\ud}{\mathrm{d}}
\newcommand{\co}[1]{\textsf{#1}}
\DeclareMathOperator*{\Tr}{Tr}

{\theoremstyle{break} \theorembodyfont{\rmfamily}

}

\begin{document}
 \title{Distribution of {\bfseries\boldmath $G$}--concurrence of random pure states}
 \medskip

 \author{Valerio Cappellini$^1$, Hans-J{\"u}rgen Sommers$^2$ and
Karol \.{Z}yczkowski$^{1,3}$}

 \affiliation{$^1$Centrum Fizyki Teoretycznej, Polska Akademia Nauk,
Al. Lotnik{\'o}w 32/44, 02-668 Warszawa, Poland}

\affiliation{$^2$Fachbereich Physik, Universit\"{a}t Duisburg-Essen,
Campus Duisburg, 47048 Duisburg, Germany}

\affiliation {$^3$Instytut Fizyki im. Smoluchowskiego, Uniwersytet
Jagiello{\'n}ski, ul. Reymonta 4, 30-059 Krak{\'o}w, Poland}

 \smallskip
 \date{November 23, 2006}

\begin{abstract}
Average entanglement of random pure states of an $N \times N$
composite system is analyzed. We compute the average value of the
determinant $D$ of the reduced state, which forms an entanglement
monotone. Calculating higher moments of the determinant we
characterize the probability distribution $P(D)$. Similar results
are obtained for the rescaled $N^{\text{th}}$ root of the
determinant, called $G$--concurrence. We show that in the limit
$N\to\infty$ this quantity becomes concentrated at a single point
$G_{\star}=1/{\textrm{e}}$. The position of the concentration point
changes if one consider an arbitrary $N\times K$ bipartite system,
in the joint limit $N,K\to\infty$, $K/N$ fixed.
\end{abstract}


\maketitle
\begin{center}
 {\small e-mail: valerio@cft.edu.pl \ \ \ sommers@theo-phys.uni-essen.de \ \ \  karol@cft.edu.pl}
\end{center}
\section{Introduction} 

Designing protocols of quantum information processing one usually
deals with some particular initial states. One is then interested in
describing the evolution of such a concrete quantum state and its
properties in time. For instance, one studies the time dependence of
the degree of quantum entanglement, which characterizes the
non--classical correlations between subsystems and is treated as a
crucial resource in the theory of quantum information~\cite{NC00}.

As a reference point one may compare the degree of entanglement of
the analyzed state with analogous properties of a typical, random
state. Such random states are also of a direct physical interest
since they arise under the action of a typical quantum chaotic
system -- see e.g.~\cite{Ha90}. In this work we investigate mean
values of certain measures of quantum entanglement, averaged over
the entire space of pure states of a Hilbert space of a given size.

There exist several measures of quantum entanglement which do not
increase under local operations and satisfy the required properties
listed in~\cite{VP98,Ho01}, but it is hardly possible to single out
the ``best'' universal quantity. On the contrary, different
entanglement measures occurred to be optimal for various tasks, so
it is likely we will have to learn to live with quite a few of
them~\cite{PV05,BZ06}.

The measures of quantum entanglement for a pure state of a bipartite
system, $|\psi\rangle\in{\cal H}={\cal H}_A\otimes{\cal H}_B$, rely
on its Schmidt coefficients~\cite{Pe93} equivalent to the spectrum
$\vec \Lambda$ of the reduced system, $\rho={\rm Tr}_B(|\psi\rangle
\langle \psi |)$. By construction the sum of all Schmidt
coefficients equals unity, $\sum_{i=1}^N \Lambda_i=1$, so just
$\pt{N-1}$ of them are independent. To quantify entanglement of a
pure state one uses entanglement monotones~\cite{Vi00}, defined as
quantities which do not increase under \co{L}ocal \co{O}perations
and \co{C}lassical \co{C}ommunication (the so--called \co{LOCC}
operations). Entanglement of a pure state of a $N \times N$ system
is therefore completely described by a suitable set of $\pt{N-1}$
independent entanglement monotones.

It is convenient to work with the ordered set of coefficients,
$\Lambda_1 \geq \Lambda_2 \geq \dots \geq \Lambda_N \geq 0$. The
first example of such a set of entanglement monotones found by Vidal
consists of sums of $k$ largest coefficients, $E_k:=\sum_{i=1}^k
\Lambda_i$ with $k=1,\dots, N-1$~\cite{Vi00}. Alternatively, one can
use R{\'e}nyi entropies of $\pt{N-1}$ different orders. Another set
of monotones may be constructed out of symmetric polynomials  of the
Schmidt coefficients of order $k=2,\ldots,N$~\cite{SZK02},
\begin{eqnarray}
\nonumber
\tau_2 &=&  \sum_{k=1}^N \sum_{l=k+1}^N \Lambda_k\Lambda_l, \nonumber \\
\tau_3 &=&  \sum_{k=1}^N \sum_{l=k+1}^N \sum_{m=l+1}^N
\Lambda_k\Lambda_l\Lambda_m,  \nonumber \\
\vdots & & \vdots\nonumber \\
\tau_N &=&  \prod_{k=1}^N\Lambda_k \ \cdot\nonumber
\end{eqnarray}
For large $N$ these polynomials become small, so it is of advantage
to consider cognate quantities,  $\tau_k'=(\tau_k)^{1/N}$. Gour
noted that taking the $N$-th root of the polynomials does not spoil
the monotonicity and proposed to used normalized quantities
$\tau_k'$ as alternative measures of quantum
entanglement~\cite{Gi05}. In particular he found unique properties
of the last polynomial $\tau_N$, equal to the determinant of the
reduced matrix, $D={\rm det}\;\rho$. Its rescaled $N$--th root,
\begin{equation}
G:=N D^{1/N}\quad,
\label{G_conc}%
\end{equation}%
proportional to the geometric mean of all Schmidt coefficients, was
called $G$--concurrence in~\cite{Gi05}, where its operational
interpretation as a type of entanglement capacity was suggested.
This quantity  extended by the convex roof construction for mixed
states, played a crucial role in demonstration of an asymmetry of
quantum correlations~\cite{HHH05} and was used to characterize the
entanglement of assistance~\cite{Gi05b}.

The aim of this work is to compute mean values and to describe
probability distributions for the determinant $D$ and its root $G$
of random pure states of a bipartite system, generated with respect
to the natural, unitary invariant measure on the space of pure
states, also called \co{F}ubini--\co{S}tudy (\co{FS}) measure.
Our analysis is performed for a bipartite system
of an arbitrary size $N$,
and in particular we treat in  detail the interesting
limiting case $N\to\infty$.
Although our study directly concerns bipartite systems,
one may infer some statements valid also in the general case
 of multipartite systems.

The paper is organized as follows. In section II we review a concept
of a random pure state and describe certain probability measures in
this set. Average values of the $G$--concurrence are computed in
section III, while the subsequent section concerns with probability
distribution of this measure of quantum entanglement. The paper is
concluded with some final remarks while the discussion of the
asymptotics of probability distributions is postponed to an
appendix.

\section{Random pure states and induced measures} 
\label{sezione2} Consider a pure state of a bipartite $N \times K$
system represented in a product basis
\begin{equation}
|\psi\rangle = \sum_{i=1}^N \sum_{j=1}^K A_{ij} |i\rangle \otimes
|j\rangle \ . %
\nonumber
\end{equation}
The Schmidt coefficients $\Lambda_i$ coincide with the eigenvalues
of a positive matrix $\rho_N=AA^{\dagger}$, equal to the density
matrix obtained by a partial trace on the $K$--dimensional space.
The matrix $A$ needs not to be Hermitian, the only constraint is the
trace condition,
 $\Tr AA^{\dagger}=1$. Furthermore,
the natural unitarily invariant measure on the space of pure states
corresponds to taking $A$ as a matrix from the Ginibre
ensemble~\cite{ZS01}. Thus our problem consists in analyzing the
distribution of determinants of random Wishart matrices
$AA^{\dagger}$ normalized by fixing its trace. Schmidt
coefficients's distributions are given by~\cite{LS88}
\begin{subequations}
\label{gen_meas}
\begin{equation}
P^{(\beta)}_{N,K}(\Lambda_1,\ldots,\Lambda_N)\ = \
B^{(\beta)}_{N,K}\; \delta\pt{1-\sum_{i}
\Lambda_i}\prod_i\Lambda_i^{\pq{\beta(K-N)+\beta-2}/2}
\theta(\Lambda_i)\prod_{i<j}|\Lambda_i-\Lambda_j|^{\beta}%
\
,
\end{equation}
in which the cases of real or complex $A$ are distinguished by the
\emph{repulsion exponent} $\beta$~\cite{Me91} being equal $1$,
respectively $2$ and the normalization $B^{(\beta)}_{N,K}$
reads~\cite{ZS01}
\begin{equation}
 B^{(\beta)}_{N,K} :=  \frac{\Gamma(KN\beta/2) {\pq{\Gamma(1+\beta/2)}}^N}
{\prod_{j=0}^{N-1} \Gamma\pq{(K-j)\beta/2}
\Gamma\pq{1+(N-j)\beta/2}} \ .\label{B_coeff}
\end{equation}
\end{subequations}
Formulae~\eqref{gen_meas} describe a family of probability measures
in the simplex of eigenvalues of a density matrix of size $N$. The
integer number $K$, determining the size of the ancilla, can be
treated as a free parameter.

\bigskip

Another important probability measure in the space of mixed quantum
states is induced by the Euclidean geometry and the
\co{H}ilbert--\co{S}chmidt (\co{HS}) distance. Assuming that each
ball
of a certain radius contains the same volume, one arrives at the
\co{HS} measure~\cite{ZS03}
\begin{subequations}
\label{consths}
\begin{equation}
P^{(\beta)}_{\text{\co{HS}}}(\Lambda_1,\ldots,\Lambda_N)\ :=  \
H_{N}^{(\beta)} \delta\pt{\sum_{i=1}^N \Lambda_i -1}\prod_{i=1}^N
\:\theta(\Lambda_i)\prod_{i<j}
|\Lambda_i-\Lambda_j|^{\beta} %
\ , \label{consths_a}
\end{equation}
where the parameter $\beta$  distinguishes as before between the real
and the complex cases. The above normalization constant
$H_N^{(\beta)}$ reads
\begin{equation}
\frac{1}{H_N^{(\beta)} }:= \frac{1}{\Gamma[N +\beta N(N-1)/2]}
\prod_{j=1}^N \Biggl[ \frac{ \Gamma(1+j\beta/2) \Gamma[1
+(j-1)\beta/2]}
     { \Gamma(1+\beta/2)} \Biggr]\quad \cdot\label{consths2}
\end{equation}
\end{subequations}
We observe that the distribution~\eqref{consths}, normalization
constants included, can be recasted into the form~\eqref{gen_meas},
provided that we choose $K=N-1+2/\beta$, that is
\begin{equation}
K=
\begin{cases}
N & \text{for complex\ } \rho_N,\ (\text{with}\ \beta=2)\\
N+1 & \text{for real\ } \rho_N,\ (\text{with}\ \beta=1)
\end{cases}\ \cdot
\label{fixK}
\end{equation}
Using this observation, one can get a useful procedure for generating
random density matrices distributed according to the
\co{HS}--measure taking normalized Wishart matrices $AA^{\dagger}$,
with $A$ belonging to the Ginibre ensemble of Hermitian matrices of
appropriate dimension.

\bigskip

Aiming to derive the averaged moments needed in
Section~\ref{ttrree}, it is convenient to change variable
in~\eqref{gen_meas} by putting $K=2\alpha/\beta+N-1$ and obtaining:
\begin{subequations}
\label{constab}
\begin{equation}
P^{(\alpha,\beta)}_{N}(\Lambda_1,\ldots,\Lambda_N) :=  \
C_{N}^{(\alpha,\beta)} \delta\pt{\sum_{i=1}^N \Lambda_i
-1}\prod_{i=1}^N \Lambda_i^{\alpha-1} \:\theta(\Lambda_i)\prod_{i<j}
|\Lambda_i-\Lambda_j|^{\beta} \label{constab_uno} \ ,%
\end{equation}
with
\begin{equation}
\frac{1}{C_N^{(\alpha,\beta)} }:= \frac{1}{\Gamma[\alpha N +\beta
N(N-1)/2]} \prod_{j=1}^N \Biggl[ \frac{ \Gamma(1+j\beta/2)
\Gamma[\alpha +(j-1)\beta/2]}
     { \Gamma(1+\beta/2)} \Biggr]\quad \cdot\label{constab2}
\end{equation}
\end{subequations}
In the above formula the real variable $\alpha$ can be used as a
free parameter instead of the integer $K$.

\section{Average moments of {\bfseries\boldmath $G$}--concurrence}
\label{ttrree}

In this Section we are going to compute averages over
an ensemble of random density matrices distributed according to the
\co{HS}--measure, which is induced by the Euclidean geometry. This
 corresponds to fixing the size $K$ of the ancilla
according to~\eqref{fixK}, depending on whether the real or the
complex case is concerned.

\begin{figure}[h]
\begin{center}
\includegraphics[width=\textwidth]{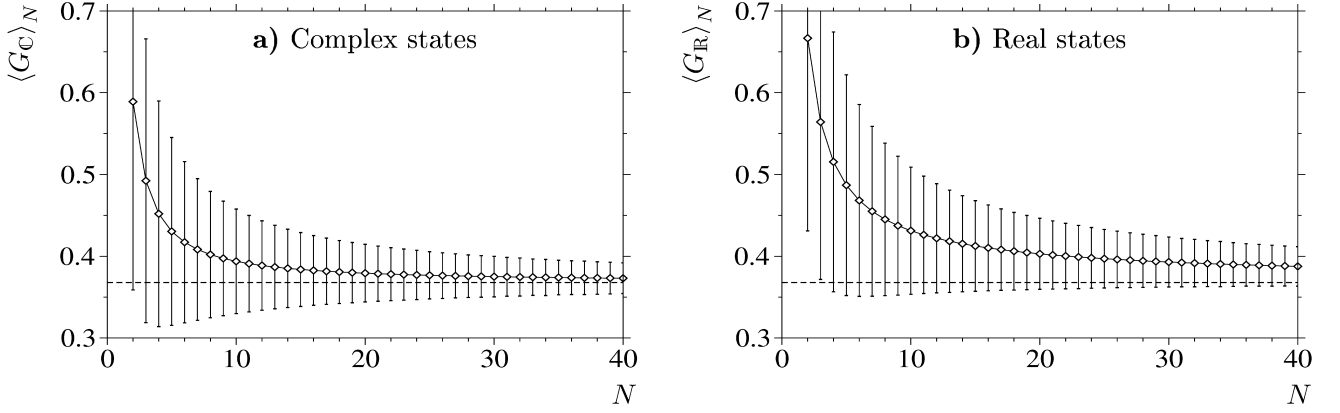}
\caption[]{Average of $G$--concurrence for {\bfseries (a)} complex
and {\bfseries (b)} real random
 mixed states of a $N\times \pt{N+2-\beta}$ system
distributed according to the \co{HS} measure. The average is
computed by means of equation~\eqref{G^M}; error bars represent the
variance of $P_N^{(\beta)} (G)$. Dashed line represent the asymptote
$G_{\star}=1/\textrm{e}$, whose explanation is given in
Section~\ref{asymptote}.} %
\label{D_mean}%
\end{center}
\end{figure}

Denoting the eigenvalues of the density matrix $\rho_N$ by
$\left\{\Lambda_j\right\}$, the moments of the determinants
$D(\Lambda_1,\ldots,\Lambda_N)=\prod_{j=1}^N \Lambda_j$ read
\begin{equation}
\langle D^M_{(\beta)} \rangle_N:=\int_{-\infty}^{\infty}
\ud\Lambda_1 \cdots \int_{-\infty}^{\infty} \ud\Lambda_N
\;D^M(\Lambda_1,\ldots,\Lambda_N)
\;P^{(\beta)}_{\text{\co{HS}}}(\Lambda_1,\ldots,\Lambda_N) \
\cdot\label{mmm}
\end{equation}
The product of Heaviside step functions, present in the
definition~\eqref{consths_a} of $P^{(\beta)}_{\text{\co{HS}}}$,
allows us to extend the domain of integration on the entire axis.
The integrand of~\eqref{mmm} coincides with the factor
present in the right hand side of equation~\eqref{constab_uno},
provided that the parameter $\alpha$
is set there to $1+M$.
 Using this the integral~\eqref{mmm} can be
computed from~\eqref{constab2}, and reads
 \begin{equation}
\begin{cases}
\langle D^M_{\mathds{C}} \rangle_N& =\frac{C_N^{\left(1 ,2\right)}}{
C_N^{\left(1+M ,2\right)}} = \frac{\Gamma\left(N^2\right)}{
\Gamma\left(MN+N^2\right)} \prod_{j=1}^N
\frac{\Gamma\left(M+j\right)}{
\Gamma\left(j\right)}\\
\langle D^M_{\mathds{R}} \rangle_N& =\frac{C_N^{\left(1 ,1\right)}}{
C_N^{\left(1+M ,1\right)}} =
\frac{\Gamma\left(\frac{N^2+N}{2}\right)}{
\Gamma\left(MN+\frac{N^2+N}{2}\right)} \prod_{j=1}^N
\frac{\Gamma\left(M+\frac{j+1}{2}\right)}{
\Gamma\left(\frac{j+1}{2}\right)}
\end{cases}
\label{D^M}\quad\cdot
\end{equation}
For sake of clarity, from now on the sub-- and super--script
${(\beta=2)}$ and ${(\beta=1)}$ will be often replaced by
${\mathds{C}}$, respectively ${\mathds{R}}$. Making use of
equation~\eqref{G_conc}, one obtains the moments of the
$G$--concurrence by imposing $\alpha=1+M/N$ in the ratios
${C_N^{\left(1 ,\beta\right)}/C_N^{\left(\alpha ,\beta\right)}}$,
rescaled  by a factor $N^M$. Thus we get now
\begin{equation}
\begin{cases}
\langle G^M_{\mathds{C}} \rangle_N& = N^M \frac{C_N^{\left(1
,2\right)}}{ C_N^{\left(1+M/N ,2\right)}} = N^M
\frac{\Gamma\left(N^2\right)}{ \Gamma\left(M+N^2\right)}
\prod_{j=1}^N \frac{\Gamma\left(\frac{M}{N}+j\right)}{
\Gamma\left(j\right)}\\
\langle G^M_{\mathds{R}} \rangle_N& = N^M \frac{C_N^{\left(1
,1\right)}}{ C_N^{\left(1+M/N ,1\right)}} = N^M
\frac{\Gamma\left(\frac{N^2+N}{2}\right)}{
\Gamma\left(M+\frac{N^2+N}{2}\right)} \prod_{j=1}^N
\frac{\Gamma\left(\frac{M}{N}+\frac{j+1}{2}\right)}{
\Gamma\left(\frac{j+1}{2}\right)}
\end{cases}
\label{G^M}%
\qquad ;
\end{equation}
in FIG.~\ref{D_mean} the mean values ${\langle
G_{(\beta)}^{\phantom{2}} \rangle}_N^{\phantom{2}}$ and variance
$\sigma^2_N={\langle G^2_{(\beta)} \rangle}_N^{\phantom{2}}-{\langle
G_{(\beta)}^{\phantom{2}} \rangle}_N^2$ are represented as a
function of $N$ for both complex and real cases.

%
\section{Probability distribution {\bfseries\boldmath $P_N^{(\beta)}(G)$}}
\begin{figure}[h]
\begin{center}
\includegraphics[width=\textwidth]{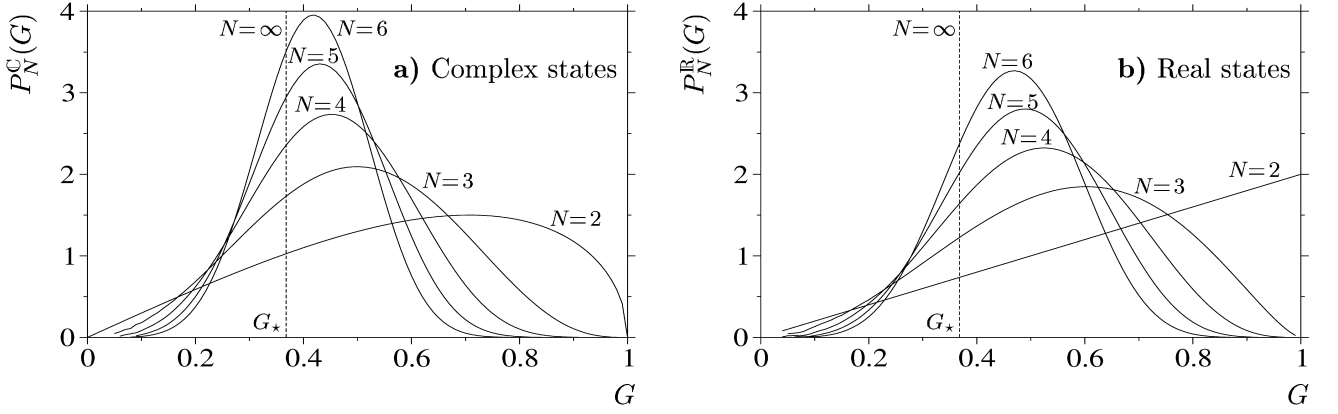}
\caption[]{$G$--concurrence's distributions $P_N^{(\beta)} (G)$ are
compared for different $N$ in the case of {\bfseries (a)} complex
and {\bfseries (b)} real random pure states. The distributions are
obtained by performing numerically~\cite{DALM} the inverse Laplace
transform of equation~\eqref{P_N (D)}. Dashed vertical line centered
in $G_{\star}=1/\textrm{e}$ denotes the position of the Dirac--delta
corresponding to $P_{N\to\infty}^{(\beta)} (G)$, as it is shown in
section~\ref{asymptote}.} %
\label{LITtog}%
\end{center}
\end{figure}

This Section is devoted to the study of probability distributions.
We shall start with the simplest problem of determining the
distribution of the determinant $D$ of a $2\times 2$ density matrix
$\rho_2$ distributed according to the (\co{HS})--measure. In this
case an explicit solution is easily obtained by integrating the
Dirac delta $\delta(D-\Lambda_1\Lambda_2)$ over the distribution
$P^{(\beta)}_{\text{\co{HS}}}(\Lambda_1,\Lambda_2)$
of~\eqref{constab}, that is
\begin{equation}
P_2^{(\beta)}(D):= C_{2}^{(1,\beta)}\int_{-\infty}^{\infty}
\ud\Lambda_1 \int_{-\infty}^{\infty} \ud\Lambda_2
\,\delta(\Lambda_1+\Lambda_2-1)\,\theta(\Lambda_1)\,\theta(\Lambda_2)
\,|\Lambda_1-\Lambda_2|^{\beta} \,\delta(D-\Lambda_1\Lambda_2)
\nonumber
\ \cdot
\end{equation}
It is a very simple distribution since
$(\Lambda_1-\Lambda_2)^2=(\Lambda_1+\Lambda_2)^2 -4D=1-4D$. Thus
\begin{equation}
\begin{cases}
P_2^{\mathds{C}}(D) & = 6 \:\sqrt{1-4D} \\
P_2^{\mathds{R}}(D) & = 4
\end{cases}\qquad,\qquad D\in\left[0,\frac{1}{4}\right]
\label{P2(D)1}%
\ \cdot
\end{equation}
The $G$--concurrence distribution $P_2^{(\beta)}(G)$ can be computed
either by integrating $\delta(G-2\sqrt{\Lambda_1\Lambda_2})$ over
$P^{(\beta)}_{\text{\co{HS}}}(\Lambda_1,\Lambda_2)$, or simply using
the latter result~\eqref{P2(D)1} together with
$P_2^{(\beta)}(G)\:\ud G = P_2^{(\beta)}(D)\:\ud D$; in both cases
(see FIG.~\ref{LITtog})
\begin{equation}
\begin{cases}
P_2^{\mathds{C}}(G) & = 3 \:G \:\sqrt{1-G^2} \\
P_2^{\mathds{R}}(G) & = 2 \:G
\end{cases}\qquad,\qquad G\in\left[0,1\right]
\label{P2(D)11}%
\ \cdot
\end{equation}
Note that, due to $\Lambda_1+\Lambda_2=1$, (only) for the case $N=2$
the $G$--concurrence given by~\eqref{G_conc} reduces to the standard
concurrence~\cite{Wo98}, $C=\sqrt{2\pt{1-\Tr\rho_2^2}}$. Thus
formula~\eqref{P2(D)11} for the complex case coincides with the
distribution of concurrence $P(C)$ obtained in~\cite{ZS01}.

\smallskip

For higher $N$ we will construct the distribution $P_N^{(\beta)}(D)$
from all moments $\langle D_{(\beta)}^M \rangle_N$ given by
equation~\eqref{D^M}; indeed
\begin{equation}
\langle D^M_{(\beta)} \rangle_N=  \int_0^1\ud D\ D^M\ P_N^{(\beta)}
(D) = \int_0^{\infty}\ud x\
\textrm{e}^{-x(1+M)}P_N^{(\beta)}(\textrm{e}^{-x}) \
,%
\nonumber
\end{equation}
with $D={\rm e}^{-x}$ and $\ud D=-\ud x\ {\rm e}^{-x}$, and so we
can obtain $P_N (D)$ by inverse Laplace transform or inverse Mellin
transform as integral along the imaginary $M$--axis:
\begin{equation}
 P_N^{(\beta)} (D)=\int_{-i \infty}^{+i \infty} \frac{\ud M }{ 2\pi i}\ D^{-(1+M)}\
\langle D_{(\beta)}^M \rangle_N %
\label{P_N (D)}%
\end{equation}
Although equations~\eqref{P_N (D)} and~\eqref{D^M} allow us to
compute the $P_N^{(\beta)} (D)$ probabilities, the cognate
quantities $P_N^{(\beta)}(G)$ can be determined as well by using
\[
P_N^{(\beta)}(G)\:\ud G = P_N^{(\beta)}(D)\:\ud D\ ;
\]
by taking from~\eqref{G_conc} the explicit expression for $\ud D/\ud
G$, one can indeed get the simple expression
\begin{equation}
G\cdot P_N^{(\beta)} (G)=N\cdot D\cdot P_N^{(\beta)} (D)\
\cdot%
\label{relation}%
\end{equation}
From now on formulae and figures will be given indifferently for
both $G$ and $D$ distribution, being clear their mutual relation. In
particular the $D$--distribution is more indicated in showing
details of calculation, for its simpler form, whereas the
$G$--distribution better shows features in the pictures, for its
domain being independent of $N$.
Important is the asymptotic behavior of the Gamma function for large
argument (Stirling's formula)
\begin{equation}
   \Gamma(z) = z^{(z-1/2)} \sqrt{2\pi} \:\textrm{e}^{-z} \pq{1 + \frac{1}{12z} +
{\mathcal{O}}{\pt{\frac{1}{z}}}^2} %
\nonumber
\end{equation}
for $|z| \to \infty$ and $\arg(z) < \pi$. This implies the
asymptotic behavior of~\eqref{D^M} for large $|M|$:
 \begin{equation}
\begin{cases}
\rule[-3ex]{0pt}{5ex}\langle D^M_{\mathds{C}} \rangle_N& \simeq
D^{\text{\co{S}}}_{\mathds{C}}\pt{M,N} :=A_N^{\mathds{C}}\cdot\frac{\textrm{e}^{-MN\log N}}{M^{(N^2-1)/2}}\\
\rule[-2ex]{0pt}{5ex}\langle D^M_{\mathds{R}} \rangle_N& \simeq
D^{\text{\co{S}}}_{\mathds{R}}\pt{M,N} :=
A_N^{\mathds{R}}\cdot\frac{\textrm{e}^{-MN\log N}}{M^{(N^2+N-2)/4}}
\end{cases}
\quad,\quad\text{with}\quad
\begin{cases}
\rule[-3ex]{0pt}{5ex} A_N^{\mathds{C}}:=
 \frac{(2\pi)^{(N-1)/2 }\Gamma(N^2)}{N^{N^2-1/2} \prod_{j=1}^N \Gamma(j)}\\
\rule[-2ex]{0pt}{5ex}A_N^{\mathds{R}}:=
 \frac{(2\pi)^{(N-1)/2 }\Gamma\pq{\pt{N^2+N}/{2}}}{N^{(N^2+N-1)/2} \prod_{j=1}^N \Gamma\pq{\pt{j+1}/{2}}}
\end{cases} %
\label{momasympt}%
\quad\cdot
\end{equation}
As a consequence the integral~\eqref{P_N (D)} converges and moreover
it vanishes if $x< N\log N$ or $D> (1/N)^N$, because in that case we
can close the contour in (\ref{P_N (D)}) in the right $M$--halfplane
according to the Jordan's Lemma~\cite{Ar85}. Physically this means
that there are no density matrices with determinants greater than
the one with maximal entropy.

In the rest of this section we will give the asymptotic behavior of
distributions $P_N^{(\beta)} (D)$ for the two edges of the domain,
that is $D\to0$ and $D\to(1/N)^N$. The details of calculation,
together with the explicit $N$--dependence of all coefficients
listed here in the following, are collected in
Appendix~\ref{appendice}.

\begin{figure}[h]
\begin{center}
\includegraphics[width=\textwidth]{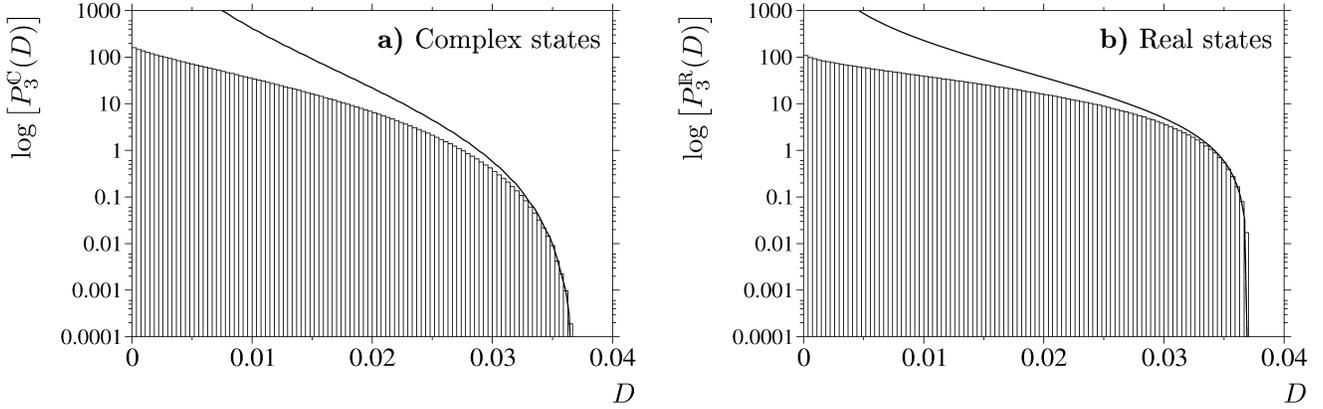}
\caption[]{In panel {\bfseries (a)} a $100$ bins histogram of $10^8$
determinants of $3\times3$ complex density matrices distributed
accordingly to the \co{HS} measure is compared with the right
asymptote given by equation~\eqref{Dasdx} (plotted in solid line).
Same analysis is depicted in panel {\bfseries (b)}, but for
$3\times3$ real density matrices.} %
\label{Asy1}%
\end{center}
\end{figure}

In particular, when very close to the completely mixed state, that
is $D\simeq(1/N)^N$, we have the result (see FIG.~\ref{Asy1})
 \begin{equation}
\begin{cases}
\rule[-3ex]{0pt}{5ex}P_N^{\mathds{C}}(D) \simeq
A_N^{\mathds{C}}\cdot\frac{(-\log D -N\log N)^{(N^2-3)/2}}{D\
{\big[\pt{N^2-3}/{2}\big]}{\displaystyle !}}\\
\rule[-2ex]{0pt}{5ex}P_N^{\mathds{R}}(D) \simeq
A_N^{\mathds{R}}\cdot\frac{(-\log D -N\log N)^{(N^2+N-6)/4}}{D\
{\big[\pt{N^2+N-6}/{4}\big]}{\displaystyle !}}
\end{cases} %
\label{Dasdx}%
\ \cdot
\end{equation}

Moreover, using~\eqref{relation} together with
 \begin{equation}
-\log D -N\log N\simeq 1-D\;N^N=1-G^N\quad, \nonumber
\end{equation}
we simply find
 \begin{equation}
\begin{cases}
\rule[-3ex]{0pt}{5ex}P_N^{\mathds{C}}(G) \simeq
{\widetilde{A}}_N^{\mathds{C}}\cdot\frac{(1-G^N)^{(N^2-3)/2}}{G}\\
\rule[-2ex]{0pt}{5ex}P_N^{\mathds{R}}(G) \simeq
{\widetilde{A}}_N^{\mathds{R}}\cdot\frac{(1-G^N)^{(N^2+N-6)/4}}{G}
\end{cases}
\quad,\quad\text{with}\quad
\begin{cases}
\rule[-3ex]{0pt}{5ex} {\widetilde{A}}_N^{\mathds{C}}:=
A_N^{\mathds{C}}\cdot
 \frac{N}{\Gamma{\big[\pt{N^2-1}/{2}\big]}}\\
\rule[-2ex]{0pt}{5ex} {\widetilde{A}}_N^{\mathds{R}}:=
A_N^{\mathds{R}}\cdot
 \frac{N}{\Gamma{\big[\pt{N^2+N-2}/{4}\big]}}
\end{cases} %
\nonumber
\quad\cdot
\end{equation}
For the other part of the spectrum, that is for very small $D$, the
probability $P_N^{\mathds{C}}(D)$ can be expanded in a power series
with some logarithmic corrections, as follows:
\begin{equation}
    P_N^{\mathds{C}}(D) \simeq Z_N^{\mathds{C}} + X_N^{\mathds{C}}\cdot D\log D + {\widetilde{X}}_N^{\mathds{C}}\cdot D
    + V_N^{\mathds{C}}\cdot D^2\pt{\log D}^2 +{\widetilde{V}}_N^{\mathds{C}}\cdot D^2\pt{\log D}+
    \widetilde{\widetilde{V}}{}_N^{\mathds{C}} \cdot D^2 + {\cal
    O}\pt{D^3(\log D)^3} %
\label{PDto0}%
\end{equation}
In particular, coefficients
$Z_N^{\mathds{C}},X_N^{\mathds{C}},{\widetilde{X}}_N^{\mathds{C}}$
are computed in appendix~\ref{appendice} for all $N\geq3$, whereas
for $V_N^{\mathds{C}},{\widetilde{V}}_N^{\mathds{C}}$ and
$\widetilde{\widetilde{V}}{}_N^{\mathds{C}}$ we limit ourself to
explicitly solve the case $N=3$ (the case $N=2$ is simply given by
formula~\eqref{P2(D)1}).
\begin{figure}[h]
\begin{center}
\includegraphics[width=\textwidth]{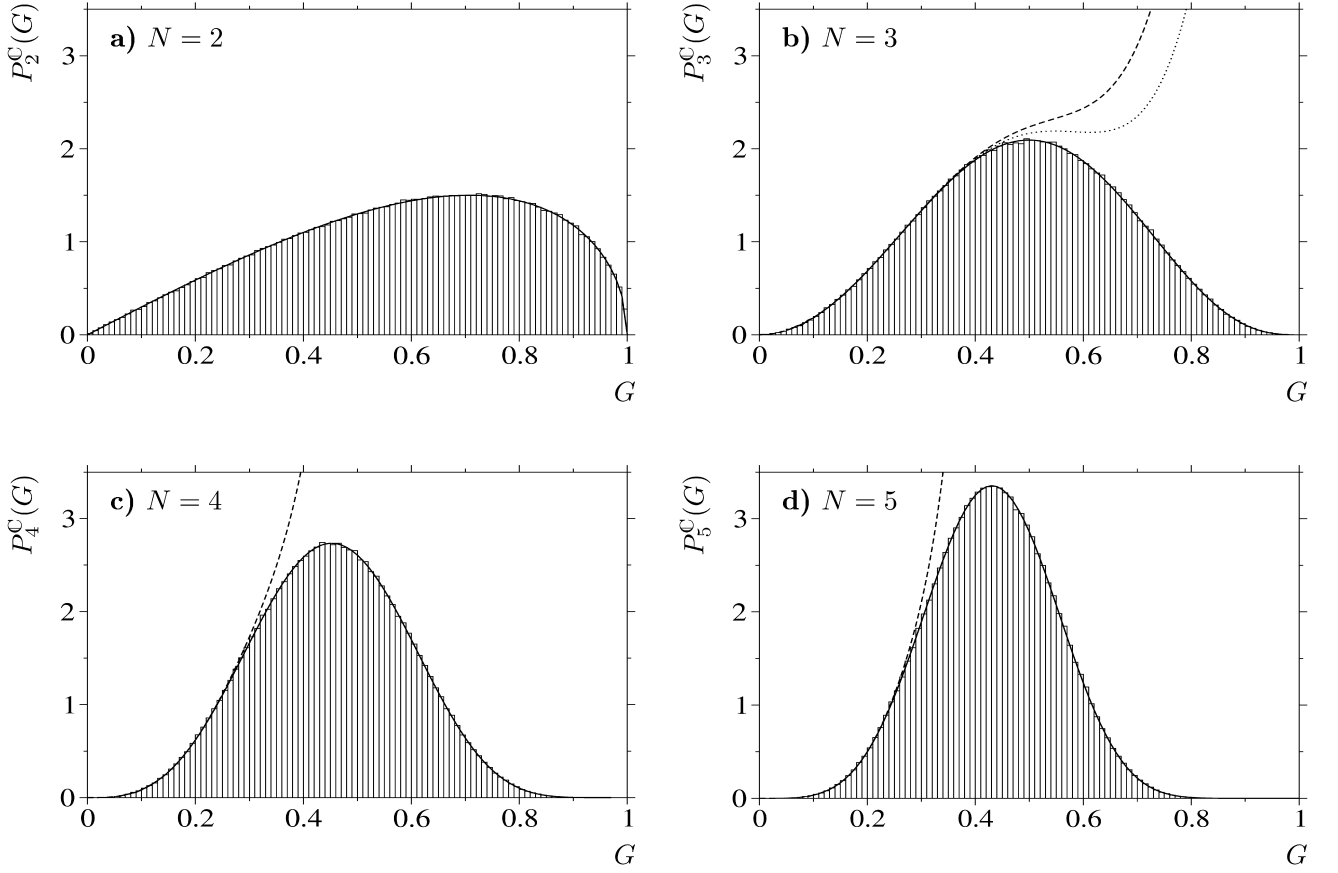}
\caption[]{In panel {\bfseries (a)}, formula~\eqref{P2(D)11} is
compared with a $100$ bins histogram of $10^6$ $G$--concurrence of
$2\times2$ complex density matrices distributed according to the
\co{HS} measure. The other panels {\bfseries (b)} {\bfseries (c)}
and {\bfseries (d)} shows histograms (for different $N$) together
with the distribution of $G$--concurrence obtained by inverse
Laplace transforming as in equation~\eqref{P_N (D)} (plotted in
solid lines). The left asymptote given by eq.~\eqref{PDto0},
computed up to ${\cal O}\pt{D}$, is also plotted in dashed line for
comparison; in panel {\bfseries (b)} we also add the contribution
given by $V_N^{\mathds{C}},{\widetilde{V}}_N^{\mathds{C}},
\widetilde{\widetilde{V}}{}_N^{\mathds{C}}$ coefficients, using a
dotted line.} %
\label{Hist}%
\end{center}
\end{figure}

The situation is similar when we do consider, in the same region of
the domain, the probability $P_N^{\mathds{R}}(D)$, corresponding to
small determinants of reduced $N\times N$ real density matrices
\co{HS}--distributed. The expansion is still a power series (plus
logarithmic corrections) but the exponents are now semi--integer,
according to the mechanism described in Appendix~\ref{appendice},
thus the probability reads:
\begin{equation}
    P_N^{\mathds{R}}(D) \simeq Z_N^{\mathds{R}} +
    Y_N^{\mathds{R}}\cdot D^{\frac{1}{2}} +
    X_N^{\mathds{R}}\cdot D\log D +{\widetilde{X}}_N^{\mathds{R}}\cdot D +
    W_N^{\mathds{R}}\cdot D^{\frac{3}{2}}\log D + {\widetilde{W}}_N^{\mathds{R}}\cdot D^{\frac{3}{2}} +
    {\cal O}\pt{D^2(\log D)^2} %
\nonumber
\end{equation}
\section{Concentration of {\bfseries\boldmath $G$}--concurrence for large system size}%
\label{asymptote}%
%
Iterating the recursion relation for the Gamma function
$\Gamma\pt{n+1}=n\:\Gamma\pt{n}$, we can recast
expression~\eqref{G^M} of the $M$--\,moment of the $G$--concurrence
of complex random pure state as
\begin{equation}
\left<G_{\mathds{C}}^M\right>_N =
\pg{\prod_{k=0}^{M-1}\frac{N}{N^2+k}}%
\pg{\left[\frac{M}{N}\;\Gamma\left(\frac{M}{N}\right)\right]^{N}}%
\pg{\prod_{k=1}^{N-1}\left(1+\frac{M}{kN}\right)^{N-k}}\quad,%
\nonumber
\end{equation}
with the asymptotics characterized with help of the Euler constant
$\gamma\approx0.\,577\;215\;665\ldots$,
\begin{align}
\prod_{k=0}^{M-1}\frac{N}{N^2+k}&\xrightarrow[\ N\to\infty\
]{}\frac{1}{N^M}\quad,
\nonumber\\%
\left[\frac{M}{N}\;\Gamma\left(\frac{M}{N}\right)\right]^{N}&\xrightarrow[\
N\to\infty\ ]{}\textrm{e}^{-\gamma M}%
\nonumber \intertext{and}
\prod_{k=1}^{N-1}\left(1+\frac{M}{kN}\right)^{N-k}&\xrightarrow[\
N\to\infty\ ]{}N^{M}\textrm{e}^{M(\gamma-1)}\quad,%
\nonumber
\intertext{so that finally}%
\left<G_{\mathds{C}}^M\right>_N &\xrightarrow[\ N\to\infty\
]{}\textrm{e}^{-M}\quad\cdot\label{cplxas}
\end{align}
For the analogue moments of $G$--concurrence of real random pure
state, some technicality requires that the sequence of odd and even
$N$ has to be analyzed separately, although it is not hard to prove
that the limit is the same. For that reason, we will simply
illustrate the case $N=2p$, $p\in\mathds{N}$, for which~\eqref{G^M}
gives
\begin{multline}
\left<G_{\mathds{R}}^M\right>_N = {\pt{\frac{2}{\sqrt{\pi}}}}^p
\pg{\prod_{k=0}^{M-1}\frac{2p}{2p^2+p+k}}%
\pg{\left[\frac{M}{2p}\;\Gamma\left(\frac{M}{2p}\right)\right]^{p}}
\pg{\left[\pt{\frac{M}{2p}+\frac{1}{2}}\;\Gamma\left(\frac{M}{2p}+\frac{1}{2}\right)\right]^{p}}\times\\
\times
\pg{\prod_{k=1}^{p-1}\left(1+\frac{M}{2pk}\right)^{p-k}}%
\pg{\prod_{k=\frac{3}{2}}^{p-\frac{1}{2}}\left(1+\frac{M}{2pk}\right)^{p-k-\frac{1}{2}}}\quad,%
\nonumber
\end{multline}
with
\begin{align}
\prod_{k=0}^{M-1}\frac{2p}{2p^2+p+k}
&\xrightarrow[\ p\to\infty\ ]{}%
{\pt{\frac{2}{2p+1}}}^M \quad,
\nonumber\\%
\left[\frac{M}{2p}\;\Gamma\left(\frac{M}{2p}\right)\right]^{p}
&\xrightarrow[\ p\to\infty\ ]{}%
\textrm{e}^{-\gamma \frac{M}{2}} \quad,
\nonumber\\%
\left[\pt{\frac{M}{2p}+\frac{1}{2}}\;\Gamma\left(\frac{M}{2p}+\frac{1}{2}\right)\right]^{p}
&\xrightarrow[\ p\to\infty\ ]{}%
\frac{\pi^{\frac{p}{2}}}{2^{\pt{p+M}}}\:\textrm{e}^{M\pt{1-\frac{\gamma}{2}}}
\quad,
\nonumber\\%
\prod_{k=1}^{p-1}\left(1+\frac{M}{2pk}\right)^{p-k}
&\xrightarrow[\ p\to\infty\ ]{}%
p^{\frac{M}{2}}\textrm{e}^{\frac{M}{2}\pt{\gamma-1}}%
\nonumber \intertext{and}
\prod_{k=\frac{3}{2}}^{p-\frac{1}{2}}\left(1+\frac{M}{2pk}\right)^{p-k-\frac{1}{2}}
&\xrightarrow[\ p\to\infty\ ]{}%
\textrm{e}^{\frac{M}{2}\pt{\gamma-1}}\textrm{e}^{-M}\:2^M{\pt{p+\frac{1}{2}}}^{\frac{M}{2}}
\quad\cdot
\nonumber%
\end{align}
Putting all factors together we arrive at the general result (compare
with~\eqref{cplxas})
\begin{equation}
G(M):=\lim_{N\to\infty}\left<G_{\pt{\beta}}^M\right>_N=
\textrm{e}^{-M}\label{genas}\qquad\cdot
\end{equation}
The above expression, valid for both $\beta\in\pg{1,2}$, is useful
to derive the limiting distribution
\begin{equation}
P^{\pt{\beta}}(G):=\lim_{N\to\infty}P_N^{\pt{\beta}}(G)\quad\cdot
\nonumber
\end{equation}
We see from~\eqref{genas} that its average is
$1/\textrm{e}=0.\,367\:879\:441\ldots$ and its variance is $0$; such
behavior can be recognized in FIG~\ref{D_mean}. Moreover, by fixing
$G=\textrm{e}^{-x}$, one can see that $G(M)$ of~\eqref{genas} is
nothing but the Laplace transform of the function
\begin{equation}
\eta(x):=\textrm{e}^{-x}P^{\pt{\beta}}(\textrm{e}^{-x})%
\nonumber
\end{equation}
so that, by inverse Laplace transforming, we obtain
\begin{equation}
\textrm{e}^{-x}P^{\pt{\beta}}(\textrm{e}^{-x})=G\:P^{\pt{\beta}}(G)=\delta(-\log(G)-1)\quad\cdot
\nonumber
\end{equation}
Rewriting the argument of the Dirac delta we finally arrive at
\begin{equation}
P^{\pt{\beta}}(G)=\delta(G-\textrm{e}^{-1})\ \cdot\label{unosue}
\end{equation}
In other words, we have shown that for large systems the
G--concurrence of random states is localized arbitrarily close to
the averaged value.

A similar concentration effect has recently been
quantified~\cite{Ha06} for bipartite $N \times K$ systems. In
particular the Von~Neumann entropy of the reduced density matrix of
the first subsystem concentrates around the entropy of the maximally
mixed state, $S\pt{\mathds{1}/N}=\log N$, if we let the dimension
$K$ of the auxiliary subsystem to go to infinity faster than $N$.
When $K=N$, so that the induced distribution coincides with the
\co{H}ilbert--\co{S}chmidt distribution, and $N\to\infty$, then
von~Neumann entropy concentrate around $\log
N-1/2$~\cite{SZ04,Ha06}. Remarkably, $G$--concurrence displays a
similar concentration effect; moreover, we are in position to prove
the convergence of its distribution to a Dirac delta centered at a
non trivial value $1/\text{e}$.

The determinants and $G$--concurrence may be also averaged in the
general case of asymmetric induced measure~\eqref{gen_meas}.
Consider an interesting case $K>N$. As for the \co{HS}--distribution
discussed in Section~\ref{sezione2} the expectation value and the
higher moments may be expressed as a ratio of normalization
constants~\eqref{B_coeff} and~\eqref{constab2}. For instance, the
moments read
\begin{equation}
{\rule[-1.5ex]{0pt}{4.5ex}\
\begin{cases}\displaystyle  \rule[-5.5ex]{0pt}{2.5ex}\langle G^M_{\mathds{C}}
\rangle_{N,K}& = N^M \frac{B^{(2)}_{N,K} }{ C_N^{(M/N+K-N+1\:,\:2)}
} =\ \displaystyle N^M \ \frac{\Gamma\left(N K\right)}{
\Gamma\left(N K+M\right)} \ \prod_{j=1}^{N}\
\frac{\Gamma\left(K\,-\,N\,+\,j\;+\;{M}/{N}\right)}{
\Gamma\left(K\,-\,N\,+\,j\right)}\\
\langle G^M_{\mathds{R}} \rangle_{N,K}\displaystyle& =N^M
\frac{B^{(1)}_{N,K} }{ C_N^{(M/N+(K-N+1)/2\:,\:1)} } =\
\displaystyle N^M\ \frac{\Gamma\left({N K}/{2}\right)}{
\Gamma\left({N K}/{2}+M\right)} \ \prod_{j=1}^{N}\
\frac{\Gamma\pq{\pt{K\,-\,N\,+\,j}/{2}+{M}/{N}}}{
\Gamma\pq{\pt{K\,-\,N\,+\,j}/{2}}}
\end{cases}\quad\cdot\ }\label{mom_on_ind}
\end{equation}
Let us now study a particular case of the induced measure, for which
we consider bipartite systems of arbitrarily large dimension, with
the only constraint that the ratio between the size $K$ of the
ancilla and the size $N$ of the principal subsystem are fixed and
greater than one. Let this ratio be expressed by the rational number
$q=\ell_2/\ell_1$, with the $\ell_1$ and $\ell_2$ integers; this
means that we are considering systems with $N=J\ell_1<K=J\ell_2$.

\noindent With the same tools used in computing~\eqref{genas}, one
can let $J$ go to infinity and obtain
\begin{align}
    \rule[-1.5ex]{0pt}{4.5ex}\
G(M)&\coloneqq\lim_{J\to\infty}\left<G_{\pt{\beta}}^M\right>_{J\ell_1,J\ell_2}=
X_q^{-M}\quad,\quad\forall\
\beta\in\pg{1,2}\quad,\label{genas2}%
\intertext{with}%
X_q&\coloneqq \;\frac{1}{\textrm{e}}\
{\pt{\frac{q}{q-1}}}^{q-1}\label{pos_q}\quad,\quad q>1\quad\cdot
\end{align}
The limiting distribution $P_{q}(G)$, can be earned as before and
reads
\begin{equation}
    \rule[-1.5ex]{0pt}{4.5ex}\
P^{(\beta)}_q(G)\coloneqq\lim_{J\to\infty}P_{J\ell_1,J\ell_2}^{(\beta)}(G)=\delta(G-X_q)\quad,
\,\ \nonumber
\end{equation}
for the complex as well as for the real case. Although the
accumulation point $X_q$ is not defined for the case $q=1$ (that is
the case in which states in the principal system are
\co{HS}--distributed), we find however $\lim_{q\rightarrow 1}
X_q=1/\textrm{e}$, confirming our previous result~\eqref{unosue}.
Moreover such values represent an infimum for $X_q$, whereas it
attains the supremum on the other part of the domain, that is for
$q\rightarrow\infty$. Such case correspond an extremely large
environment, for which $X_{\infty}=1$, that is in turn the
$G$--concurrence of the completely mixed state. Thus we find another
evidence that large environment concentrates reduced density
matrices around the maximally mixed states~\cite{Ha06}.

\section{Concluding remarks} 

The generalized $G$--concurrence is likely to be the first measure
of pure state entanglement for which one could find not only the
mean value over the set of random pure states, but also compute
explicitly all moments and describe its probability
distribution,
deriving an analytic expression in the large $N$ limit.
This offers for our work various potential applications.
On one hand, analyzing a concrete quantum state
and its entanglement we may check,
to what extend its properties are non typical.
In practice this can be done by a comparison
of its $G$--concurrence $G$ with the mean value $\langle G\rangle$,
and by comparing its deviation from the average, $|G-\langle G\rangle|$,
with the root of the variance of the distribution.

On the other hand, if one needs a quantum state of some particular
properties, one may estimate how difficult it is to obtain such a
state at random. For instance, looking for a state of a large degree
of entanglement, with concurrence greater than a given value
$\widetilde{G}$, one can make use of the derived probability
distribution by integrating it from $\widetilde{G}$ to unity in
order to evaluate the probability to generate the desired state by a
fully uncontrolled, chaotic quantum evolution.

Although in this work we have concentrated our attention on pure
states of bipartite systems, the averages obtained for the
asymmetric induced measures~\eqref{gen_meas} with $K>N$ may be
easily applied for the more general, multipartite case. Consider a
system containing $n$ qudits (particles described in a
$d$--dimensional Hilbert space). This system may be divided by an
arbitrary bi--partite splitting into $m$ and $(n-m)$ particles, and
one can study entanglement between both subsystems -- see
e.g.~\cite{KZM02}. The partial trace over $m$ qudits is equivalent
to the partial trace performed over a single ancilla of size
$K=d^{m}$, so setting size of the system $N=d^{n-m}$ one may read
out the average concurrence from eq.~\eqref{mom_on_ind}.  In
particular, if $n$ is even and we put $m=n/2+k$, then the ratio
$q=K/N$ is equal to $d^{2k}$ and in the asymptotic limit $n\to
\infty$ the concurrence concentrates around the mean~\eqref{pos_q}
which depends only on the asymmetry $k$ of the splitting.

 Our research may also be
considered as a contribution to the random matrix theory: we have
found the distribution of the determinants of random Wishart
matrices $AA^{\dagger}$, normalized by fixing their trace.
Furthermore, the analysis of the distribution of $G$-concurrence in
the limit of large system sizes provides an illustrative example of
the geometric concentration effect, since in high dimensions the
distribution of the determinant is well localized around the mean
value. This observation can also be related to the central limit
theorem applied to logarithms of the eigenvalues of a density
matrix, the sum of which is equal to the logarithm of the
determinant.

\medskip

It is a pleasure to thank P. Hayden and P. Horodecki for stimulating
discussions. This work was financed by the \co{SFB}/Transregio--12
project financed by \co{DFG}. We also acknowledge support provided
by the EU research project \co{COCOS} and the grant $1$\ \
\co{P}$03$\co{B}\ \ $042$\ \ $26$\ of Polish Ministry of Science and
Information Technology.

\appendix

\section{Coefficients of asymptotic expansions of probability}%
\label{appendice}%
\subsection{Right asymptote of {\bfseries\boldmath $P_N^{(\beta)} (D)$}: proof of
equation~{\bfseries(\ref{Dasdx})}}%
The starting point is integral~\eqref{P_N (D)}. Since all the poles
of the integrand are in the left half--plane (see it
in~\eqref{D^M}), the contour integration along the imaginary axis
can be modified into the one along the right asymptotic half--plane,
that is on a very large semicircle connecting $-i\infty$ to
$+i\infty$; this allow us to use the Stirling's formula for
replacing $\langle D^M_{\pt{\beta}} \rangle_N$ with
$D^{\text{\co{S}}}_{\pt{\beta}}\pt{M,N}$ (see
formula~\eqref{momasympt}) in the integrand of~\eqref{P_N (D)}. Of
course we made an approximation, but we know that the formula we
ended up matches the correct result ($P_N^{(\beta)} (D)=0$ for
$D>{\pt{1/N}}^N$) in the point ${\pt{1/N}}^N$, so that such
approximation would hold  close to that point. Now we observe that
$D^{\text{\co{S}}}_{\pt{\beta}}\pt{M,N}$ has poles only in $M=0$, so
that our contour of integration can be modified provided that we do
not cross the origin, and we do so obtaining
\begin{equation}
 P_N^{(\beta)} (D)=\int_{\gamma} \frac{\ud M}{2\pi i}\ D^{-(1+M)}\
D^{\text{\co{S}}}_{\pt{\beta}}\pt{M,N}=\frac{A_N^{\pt{\beta}}}{2\pi
i D}\int_{\gamma} {\ud M}\ {\textrm{e}^{M\pt{-\log D - N\log N}}
M^{-\frac{(N^2-1)}{2}}}
\nonumber
\quad,
\end{equation}
where $\gamma$ is now the contour that, starting from $-i\infty$ get
close to the negative real axis on the asymptotic left--lower
quarter--plane, winds around $\mathds{R}^{-}\cup\pg{0}$ in the
counterclockwise direction, and then approaches $+i\infty$ on the
asymptotic left--upper quarter--plane. But now we apply once more
Jordan's Lemma and we remove the asymptotic semi--circle from
$\gamma$. After rescaling $M\to-M/\varepsilon$, with the latter
defined by $\varepsilon=-\log D - N\log N$ and close to $1$, we
arrive at the well known Hankel's contour integral for the inverse
of the Gamma function ($1/\Gamma$)~\cite{AS70}, that leads
to~\eqref{Dasdx} and gives the asymptotic behavior for $ D \to
(1/N)^N$.

\subsection{Left asymptote of {\bfseries\boldmath $P_N^{\mathds{C}} (D)$}
 for complex random pure states}%

Now let us consider the behavior of $P^{\mathds{C}}_N(D)$ at the
lower edge of the spectrum $D \to 0$. In that case one can close the
integral~\eqref{P_N (D)} in the left halfplane obtaining
contributions from all the poles of the Gamma functions in $\langle
D^M_{\mathds{R}} \rangle_N$ (see~\eqref{D^M}). Such poles are
located at each of the negative integers $M=-1,-2,-3,\ldots$;
fortunately there is the factor $D^{-(1+M)}$ such that we obtain a
series in powers of $D$. Because of the multiple Gamma functions
in~\eqref{D^M}, most of the poles are degenerate and the general
feature (for an arbitrary large $N$) is that the pole in $-\ell$ is
of order $\ell$: due to this fact the $D$--powers in the expansion
get in general a logarithmic correction. The first pole at $M_1=-1$
is non degenerate and yields
\begin{equation}
    P^{\mathds{C}}_N(0)= \frac{\Gamma(N^2)}{\Gamma(N^2-N)\;\Gamma(N )}
    =Z_N^{\mathds{C}}\quad\cdot
\nonumber
\end{equation}
Including the next order--2 pole ($M_2=-2$) contribution we find the
asymptotic expansion for $D \to 0$
\begin{equation}
    P_N^{\mathds{C}}(D) \simeq Z_N^{\mathds{C}} + X_N^{\mathds{C}}\cdot D\log D + {\widetilde{X}}_N^{\mathds{C}}\cdot D
\nonumber
\end{equation}
with
\begin{equation}
\begin{cases} X_N^{\mathds{C}}&=\frac{\displaystyle\Gamma(N^2)}{\displaystyle\Gamma(N^2-2N)\;\Gamma(N)\;\Gamma(N-1)}\rule[-3ex]{0pt}{5ex}\\%
{\widetilde{X}}_N^{\mathds{C}}&={\textstyle X_N^{\mathds{C}}\Big(N+N\psi(N^2-2N)-4-2\psi(1)-(N-2)\psi(N-2)\Big)}%
\end{cases}
\label{CDto0}%
 \quad\cdot
\end{equation}
Here $\psi(x)$ is the Digamma function~\footnote{In
equation~\eqref{CDto0} we have used $ \sum_{k=1}^n
\psi\pt{k}=n\:\psi\pt{n}-n+1$.}, or polygamma function of order $0$,
with
\begin{equation}
      \psi(1)=-\gamma,\qquad \psi(n)=-\gamma  +
\sum_{k=1}^{n-1}\frac{1}{k},\text{\ for\ } n>1\ .\ %
\label{Digamma}%
\end{equation}
Note that the Euler constant $\gamma$ cancels everywhere. By adding
the next order--3 pole ($M_3=-3$) contribution one gets in general
the terms in~\eqref{PDto0} corresponding to the
$V_N^{\mathds{C}},{\widetilde{V}}_N^{\mathds{C}}$ and
$\widetilde{\widetilde{V}}{}_N^{\mathds{C}}$ coefficients, although
the latter are in general rather complicated, involving polygamma
function of order higher than $0$. This is not the case when $N=3$,
for which a cancelation makes $M_3=-3$ a pole of order $2$, and the
coefficients read :
\begin{equation}
      V_3^{\mathds{C}}=0\qquad,\qquad{\widetilde{V}}_3^{\mathds{C}}=6\cdot7!=30\,240\qquad,
      \qquad\widetilde{\widetilde{V}}{}_3^{\mathds{C}}=9\cdot7!=45\,360\qquad\cdot\
\nonumber
\end{equation}
\subsection{Left asymptote of {\bfseries\boldmath $P_N^{\mathds{R}} (D)$} for real random pure states}%
We will apply the same reasoning of the previous case, just now
differing for the fact that, when $\beta=1$, the $\ell^{\text{th}}$
pole $M_\ell$ of the integrand of~\eqref{P_N (D)} is
$-\frac{\ell+1}{2}$; in general, for arbitrarily large $N$, its
corresponding order is given by $\left\lfloor \frac{\ell+1}{2}
\right\rfloor$, where $\left\lfloor x \right\rfloor$ means the
larger integer not exceeding $x$. In particular, the firsts two
poles $M_1=-1$ and $M_2=-\frac{3}{2}$ are non degenerate and
yield~\footnote{From now on we will often make use of the identity
$\Gamma\pt{n/2}\Gamma\pt{\pt{n+1}/2}=\sqrt{\pi}\;\Gamma\pt{n}/2^{n-1}$,
$n\in\mathds{N}^+$.}
\begin{equation}%
Z_N^{\mathds{R}}=\frac{2^{N-1}\;\Gamma\big(\frac{N^2+N}{2}\big)}{\Gamma\pt{\frac{N^2-N}{2}}\;\Gamma\pt{N}}%
\qquad\text{and}\qquad%
Y_N^{\mathds{R}}=-\sqrt{\pi}\frac{2^{N-1}\;\Gamma\big(\frac{N^2+N}{2}\big)}{\Gamma\pt{\frac{N^2-2N}{2}}\;\Gamma\pt{\frac{N+1}{2}}\;\Gamma\pt{N-1}}%
\qquad\cdot%
\nonumber
\end{equation}
Including the next two $2$--order poles contributions ($M_3=-2$ and
$M_4=-\frac{5}{2}$) we determine, for $N>3$
\begin{align}
\,&\begin{cases}X_N^{\mathds{R}}&=-\frac{2^{2N-3}\;\Gamma\big(\frac{N^2+N}{2}\big)}{\Gamma\pt{\frac{N^2-3N}{2}}\;\Gamma\pt{N}\;\Gamma\pt{N-2}}\rule[-3ex]{0pt}{5ex}\\
{\widetilde{X}}_N^{\mathds{R}}&={\textstyle
X_N^{\mathds{R}}\pg{N+N\psi\pt{\frac{N^2-3N}{2}}-8-\frac{3}{2}\:\psi\pt{\frac{1}{2}}-2\:\psi\pt{1}-\frac{N-3}{2}\:\psi\pt{\frac{N-3}{2}}-\frac{N-4}{2}\:\psi\pt{\frac{N-4}{2}}}}%
\end{cases}%
\label{PN(1)_R1}%
\quad,
\intertext{and for $N>4$}
\,&\begin{cases}W_N^{\mathds{R}}&=-\frac{\sqrt{\pi}}{3}\frac{2^{2N-3}\;\Gamma\big(\frac{N^2+N}{2}\big)}{\Gamma\pt{\frac{N^2-4N}{2}}\;\Gamma\pt{\frac{N+1}{2}}\;\Gamma\pt{N-1}\;\Gamma\pt{N-3}}\rule[-3ex]{0pt}{5ex}\\%
{\widetilde{W}}_N^{\mathds{R}}&={\textstyle
W_N^{\mathds{R}}\pg{N+N\psi\pt{\frac{N^2-4N}{2}}-\frac{35}{3}-\frac{5}{2}\:\psi\pt{\frac{1}{2}}-2\:\psi\pt{1}-\frac{N-4}{2}\:\psi\pt{\frac{N-4}{2}}-\frac{N-5}{2}\:\psi\pt{\frac{N-5}{2}}}}
\end{cases}%
\label{PN(1)_R2}%
\quad,
\end{align}
where we made use once more of the $\psi$--digamma
function~\footnote{In
equations~\mbox{(\ref{PN(1)_R1}--\ref{PN(1)_R2})} we have used $
\sum_{k=1}^n
\psi\pt{{k/2}}=\pt{n/2}\psi\pt{{n}/{2}}+\pt{\pt{n-1}/2}\psi\pt{\pt{n-1}/{2}}+\pt{1/2}\psi\pt{{1/2}}-n+2$;
moreover, the notation $\displaystyle
0\:\psi\pt{0}=\lim_{\varepsilon\to
0}\varepsilon\:\psi\pt{\varepsilon}=-1$ is understood.}
of~\eqref{Digamma}. The case $N=3$ constitutes an exception for
$X_3^{\mathds{R}}$ and $W_3^{\mathds{R}}$'s coefficients, because of
the lowering of the order of $M_3$ and $M_4$ poles; moreover, for
the latter pole, the same happens also for $N=4$. All these
coefficients need separate calculations and read
\begin{equation}
{\widetilde{X}}_3^{\mathds{R}}=12 \cdot5!%
\hspace{2ex},\hspace{2ex}%
{\widetilde{W}}_3^{\mathds{R}}=4  \cdot5!%
\hspace{2ex},\hspace{2ex}%
{\widetilde{W}}_4^{\mathds{R}}=2^8\cdot8!%
\hspace{4ex}\text{and}\hspace{4ex}%
X_3^{\mathds{R}}= W_3^{\mathds{R}}=W_4^{\mathds{R}}=0%
\hspace{2ex}\cdot \nonumber
\end{equation}

\end{document}